\documentclass[conference]{IEEEtran}
\IEEEoverridecommandlockouts
\usepackage{cite}
\usepackage{amsmath,amssymb,amsfonts}
\usepackage{amsmath,amsthm,amssymb,mathtools}

\usepackage{physics}
\usepackage{graphicx}
\usepackage{textcomp}
\usepackage{xcolor}
\usepackage[draft]{changes}
\usepackage{algorithm}
\usepackage[utf8]{inputenc}
\usepackage{algpseudocode}
\usepackage{array}
\usepackage[caption=false,font=normalsize,labelfont=sf,textfont=sf]{subfig}
\usepackage{textcomp}
\usepackage{stfloats}
\usepackage{url}
\usepackage{verbatim}
\usepackage{graphicx}
\usepackage{cite}
\def\BibTeX{{\rm B\kern-.05em{\sc i\kern-.025em b}\kern-.08em
    T\kern-.1667em\lower.7ex\hbox{E}\kern-.125emX}}
\begin{document}

\title{Measurement Strategies and Estimation Precision in Quantum Network Tomography}
% \thanks{This research was supported by SFI (Science Foundation Ireland) grant 21/US-C2C/3750 for CoQREATE (Convergent Quantum REsearch Alliance in Telecommunications), CONNECT-2 grant, and by the NSF-ERC Center for Quantum Networks grant EEC- 1941583.}
% }

\author{\IEEEauthorblockN{Athira Kalavampara Raghunadhan\textsuperscript{1},  Matheus Guedes De Andrade\textsuperscript{2}, Don Towsley\textsuperscript{2}, Indrakshi Dey\textsuperscript{3},\\   Daniel Kilper\textsuperscript{1}, Nicola Marchetti\textsuperscript{1} }
\\
\IEEEauthorblockA{\textsuperscript{1}CONNECT Research Centre, School of Engineering, Trinity College Dublin, Ireland}
\IEEEauthorblockA{\textsuperscript{2}College of Information and Computer Science, University of Massachusetts, Amherst, USA}
\IEEEauthorblockA{\textsuperscript{3}Walton Institute, South East Technological University, Waterford, Ireland}}

\maketitle
\begin{abstract}
This work investigates measurement strategies for link parameter estimation in Quantum Network Tomography (QNT), where network links are modeled as depolarizing quantum channels distributing Werner states. Three distinct measurement schemes are analyzed: local Z-basis measurements (LZM), joint Bell-state measurements (JBM), and pre-shared entanglement-assisted measurements (PEM). For each scheme, we derive the probability distributions of measurement outcomes and examine how noise in the distributed states influences estimation precision. For each scheme, we derive closed-form expressions for the Quantum Fisher Information Matrix (QFIM) and evaluate the estimation precision through the Quantum Cramér–Rao Bound (QCRB). Numerical analysis reveals that the PEM scheme achieves the lowest QCRB, offering the highest estimation accuracy, while JBM provides a favorable balance between precision and implementation complexity. The LZM method, although experimentally simpler, exhibits higher estimation error relative to the other schemes; however, it outperforms JBM in high-noise regimes for single-link estimation. We further evaluate the estimation performance on a four-node star network by comparing a JBM-only configuration with a hybrid configuration that combines JBM and LZM. When two monitors are used, the JBM-only strategy outperforms the hybrid approach across all noise regimes. However, with three monitors, it achieves a lower QCRB only in low-noise regimes with heterogeneous links. The results establish a practical basis for selecting measurement strategies in experimental quantum networks, enabling more accurate and scalable link parameter estimation under realistic noise conditions.
\end{abstract}


\section{Introduction}
Quantum networks connect distant nodes by creating shared entanglement that enable secure communication, distributed sensing \cite{r23}, and large-scale quantum computation \cite{r10}. The ability to distribute and maintain high-quality entanglement across such networks is critically dependent on the quality of the underlying quantum links. Links that can generate entangled states with high entanglement fidelity are essential for reliable entanglement swapping, quantum teleportation \cite{r9}, and adaptive routing \cite{r8} strategies in complex network topologies. This underscores the importance of precise link characterization---specifically, accurately estimating the error rates of individual links---which is the primary goal of Quantum Network Tomography (QNT). 

Initial QNT methods focused on estimating link parameters in quantum star networks using multi-party entanglement distribution and measurements at the end nodes. These studies \cite{r1,r2,r3} provided foundational algorithms for network estimation of Pauli channels, particularly bit-flip and depolarizing noise, and analyzed their efficiency using the Quantum Fisher Information Matrix (QFIM)~\cite{r21,r22}. They established the significance of combining the quantum state distribution and measurements for parameter identifiability \cite{r18,r19}. Building upon these foundations, previous work \cite{r4} introduced the concept of monitor nodes\cite{r5,r6,r7} in QNT, and investigated single-monitor placement strategies together with bipartite entanglement generation in quantum star networks. A monitor refers to a network node capable of performing end-to-end quantum measurements to estimate channel parameters.
%\cite{r24}

In practice, it is often infeasible to directly access individual network links, as some may be constrained by physical or operational factors and others may be concurrently engaged in multiple communication tasks. These limitations necessitate end-to-end estimation strategies \cite{r11,r12,r13} that characterize the quality of an entire path (which may include multiple links) to infer the parameters of the individual links it traverses.

Path-based estimation relies on the generation of entangled states over several connected links, applying intermediate operations such as entanglement swapping \cite{r20}, and then measuring the resulting end-to-end entanglement. This approach allows estimating links of the network without direct measurements by generating entangled states that depend on the link parameters of each link in the path, although it displays inherent trade-offs. Indeed, entanglements generated over longer paths in hop distance contain information for more link parameters compared to shorter paths. However, utilizing more links in a measurement path leads to more accumulated noise, which diminishes the achievable estimation precision.

This work addresses how path-based estimation can be combined with measurement operations to perform QNT, evaluating performance through the Quantum Cramér-Rao bound (QCRB). We compare three schemes for estimating link parameters: local Z-basis measurements (LZM), joint Bell state measurements (JBM), and pre-shared entanglement-assisted measurements (PEM). The LZM scheme restricts operations to single-qubit Z-basis measurements, with correlations extracted through classical post-processing. The JBM scheme uses two-qubit operations on two noisy copies of entangled states generated across the same network path. The PEM scheme leverages a pre-shared ancillary noise-free EPR pair to measure one copy of an entangled state generated through a path in the Bell basis. Consequently, the three schemes yield distinct precision regimes for link parameter estimation, and identifying their domains of optimality is essential for practical QNT implementations. We do not provide complete descriptions for link parameter estimators, although they can be obtained from the probability distribution of outcomes described in this paper.

\subsection{Summary of Contributions} 
The main contributions of this work are as follows:

\begin{itemize}
    \item \textbf{Measurement schemes.} 
    We analyze three measurement strategies for quantum network tomography: 
    (i) LZM, 
    (ii) JBM, and 
    (iii) PEM. 
    For each scheme, we derive the probability distributions of measurement outcomes and highlight their noise dependence.

    \item \textbf{Fisher information analysis.}
    We provide closed-form expressions for the QFIM corresponding to LZM and PEM, and generalize the QFIM expression for JBM derived in~\cite{r4} for stars to paths of arbitrary length. 

    \item \textbf{Numerical validation.} 
    We evaluate the measurement strategies on a single link and a four-node star network. The results demonstrate that PEM consistently achieves the highest estimation accuracy across all noise regimes, followed by JBM. Although LZM generally yields a higher QCRB, it surpasses JBM in high-noise conditions for a single link. Further evaluations on the four-node star network reveal a similar trend only when three monitors are deployed and the links are heterogeneous.
\end{itemize}

The rest of the paper is structured as follows. Section II defines the system model and monitoring framework. Section III describes the state distribution and measurement methods, derives closed-form expressions for the QFIM, and explains the QNT framework for the star network. Section IV reports the numerical evaluation results, including validation on a four-node star network. Section V concludes with a discussion of the main findings and potential future directions.

\section{System Model}
In this paper, we adopt the network model from our previous work \cite{r4}, where a quantum network is represented as a graph $G=(V,E)$, where $V$ is the set of quantum processors and \(E\) the set of quantum links. Each link in the network is modeled as a depolarizing channel \cite{r17} used to generate a \emph{Werner state} \cite{r14,r15,r16}---a mixed bipartite state that interpolates between a maximally entangled Bell state and the maximally mixed state---between its adjacent nodes. Werner states model realistic quantum links as they capture the impact of depolarizing noise through an adjustable parameter $w$, representing how channel imperfections affect the entanglement fidelity of a link-level entangled state, e.g., an entangled state generated through the link. This makes them a practical and analytically convenient choice for studying noisy quantum networks. For $w \in [0,1]$, the Werner state is
\begin{equation}
\rho(w) = w\,|\Phi^+\rangle\!\langle \Phi^+| + (1-w)\,\frac{\mathbb{I}_4}{4},
\end{equation}
where $\ket{\Phi^+} = (\ket{00} + \ket{11})/{\sqrt{2}}$ and $\mathbb{I}_4$ is the $4\times 4$ identity matrix.
When $w=1$, $\rho(w)$ is a pure Bell state, while $w=0$ corresponds to the maximally mixed state. The entanglement fidelity of state $\rho(w)$ is defined as
\begin{align}
    & f_{\Phi}(w) = \frac{1 + 3w}{4}.
\end{align}

The parameter $w$ serves as a measure of link quality based on its connection with entanglement fidelity. In a network, different links may have different parameters, and QNT seeks to estimate link parameters for all network links with path-estimation.

\subsection{Monitor Placement and Link Monitoring}

Monitors are nodes allowed to perform quantum measurements for parameter estimation. In its seminal definition, QNT considered that external network nodes (end-nodes), e.g., leaves in a star-shaped network, were the nodes capable of performing quantum measurements for estimation. In turn, intermediate nodes, e.g, the central node in a star-shaped quantum network, were only allowed to perform operations in order to establish quantum communication among end-nodes. Monitors generalize the initial definition of QNT based on end-nodes, decoupling the place of a node in the network topology from its functionalities.

The $k$-monitor placement problem refers to: (1) specifying $k$ monitors in $V$, and (2) defining how monitors must be used to estimate link parameters in order to optimize a given objective function. This work focuses on (2), investigating how different quantum state distribution and measurement schemes affect the quality of QNT estimators when a monitor placement is fixed. It serves as an initial step towards addressing the monitor placement problem in more generic scenarios than the initial formulation considered in~\cite{r4}.

Let $e = (v_0, v_1) \in E$ denote a link in $G$ characterized by the Werner parameter $w$. A monitor $v \in V$ monitors link $e \in E$ if quantum measurements for the estimation of $w$ are performed in $v$, i.e., there is a path $P$ used to generate an entangled state to estimate $w$ containing $v$. Moreover, we say that link $e$ is directly monitored by $v$ if $v$ monitors $e$, and $e$ is the only link in the path used to generate an entangled state for estimation. A link is indirectly monitored by $v$ if $v$ monitors $e$ and the path used for entanglement generation contains more than one link.

% \subsection{State Distribution and Measurement schemes}
\section{Methods}
In the QNT formulation considered in this paper, a quantum network is probed by generating entangled states through paths, followed by measurements to characterize its behavior. This work focuses on two types of entanglement generation over a path for network estimation: linear generation and cyclic generation.

Let $P = (v_{0}, v_1, \ldots, v_n)$ be a path interconnecting nodes $v_0$ and $v_{n}$. Assume that link $(v_i, v_{i + 1})$ is characterized by parameter $w_{i}$, for $i = 0, \ldots, n - 1$. In linear generation, each link in $P$ is used to generate one copy of the Werner state $\rho(w_i)$, and each intermediate node $v_i$ performs a Bell State Measurement (BSM) to generate a copy of the end-to-end Werner state $\rho(w)$ between nodes $v_0$ and $v_n$, where $w = \prod_{i = 0}^{n - 1} w_i$. In cyclic generation, linear generation is used twice to generate two copies of the end-to-end Werner state $\rho(w)$, i.e., the state $\rho(w) \otimes \rho(w)$, followed by a BSM in node $v_{n}$ to generate the state $\rho(w^{2})$ in node $v_0$. 

\begin{figure*}
\centering
\includegraphics[scale=0.40]{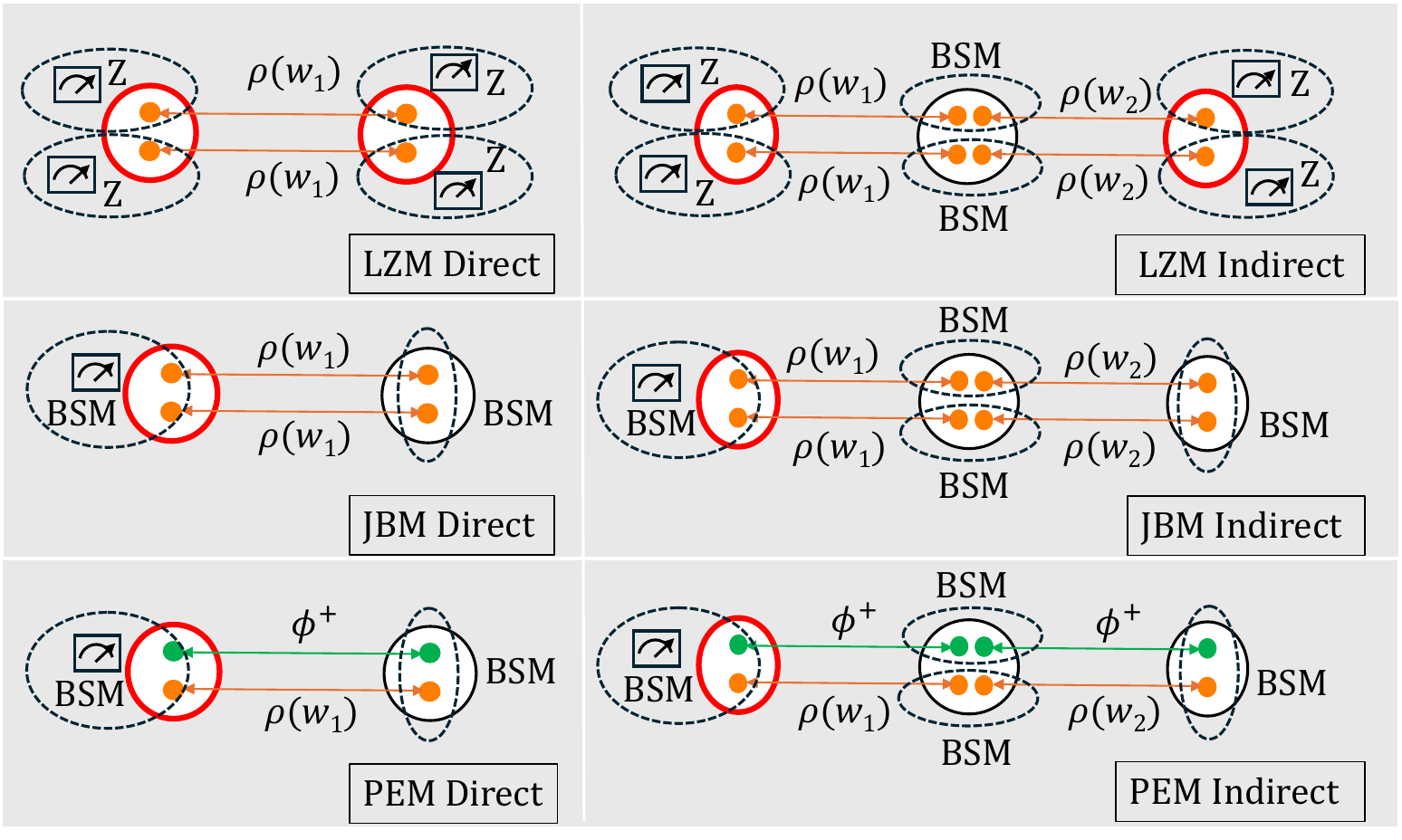}
\caption {State distribution and measurement schemes. Red-outlined nodes denote monitors, and Werner states $\rho(w_i)$ are distributed along each link. The perfect Bell state $\phi^{+}$ is distributed in PEM. In JBM and PEM, links connected to monitor are measured directly (JBM Direct and PEM Direct), while others are estimated indirectly (JBM Indirect and PEM Indirect). LZM requires monitors at both end nodes of a link or path to perform LZM Direct and LZM Indirect measurements.}
\label{0}
\end{figure*}

Fig. \ref{0} illustrates the three state distribution and measurement schemes utilized in this work that are derived from linear and cyclic generation and are explained in the following.

\subsubsection{Local Z-basis Measurements (LZM)}
Let $v_0 = A$ and $v_n = B$ be the two monitors. In this scheme, linear generation is used to create $\rho(w)$ between $A$ and $B$. Each monitor then performs local measurements in the computational ($Z$) basis on its respective qubits of $\rho(w)$. The measurement outcomes, $\{00, 01, 10, 11\}$, obtained from the two monitors are then classically correlated to estimate the Werner parameter $w$.

For the distributed Werner state, the $Z$-basis measurement outcomes follow the probability distribution
\begin{equation}
p_{00} = p_{11} = \frac{1 + w}{4}, \quad
p_{01} = p_{10} = \frac{1 - w}{4}.
\end{equation}
This linear dependence on $w$ reflects the direct influence of channel noise on the measurement statistics.

\subsubsection{Joint Bell-State Measurements (JBM)}

Assume that $A$ is a monitor. In this scheme, cyclic generation is used to generate the Werner state $\rho(w^{2})$ in node $A$. Node $A$ then perform a BSM on $\rho(w^{2})$, which is a local operation since both qubits reside in $A$. The BSM outcomes, $\{\Phi^+, \Phi^-, \Psi^+, \Psi^-\}$, obtained at node $A$ are used to estimate the Werner parameter $w$. The probability distribution of the resulting outcomes can be expressed as follows:
\begin{align}
p_{\Phi^+} = \frac{1+3w^2}{4},\quad 
p_{\Phi^-} = p_{\Psi^+} = p_{\Psi^-} = \frac{1-w^2}{4}.   
\end{align}

Werner parameters for links in a path compose multiplicatively during entanglement swapping. The quadratic dependence highlights that each  link in a path is used twice to generate the measured state in JBM.

\subsubsection{Pre-shared Entanglement-assisted Measurements (PEM)}
In this scheme, linear generation is used to distribute one copy of the Werner state $\rho(w)$ between $A$ and $B$. Assuming that $A$ and $B$ share a perfect Bell pair $\ket{\Phi^+}_{A_0B_0}$, $B$ performs a BSM in its qubits, i.e., the qubit storing its share of $\rho(w)$ and $B_0$, preparing the state $\rho(w)$ in $A$. Then, Node $A$ performs a BSM in $\rho(w)$, which is a local operation since both qubits reside in $A$.

Since only the Werner state carries noise while the pre-shared EPR pair is noiseless, the resulting joint statistics depend linearly on $w$:
\begin{align}
p_{\Phi^+}= \frac{1+3w}{4}, \quad
p_{\Phi^-}=p_{\Psi^+}=p_{\Psi^-} = \frac{1-w}{4}.  
\end{align}
Implementing PEM requires additional pre-shared entanglement in the form of a perfect Bell state, which is non-trivial to generate, making this scheme resource-intensive.

\subsection{Quantum Fisher Information Matrix Analysis}

We quantify the performance of each state distribution and measurement scheme using the QFIM, which captures the sensitivity of the measurement outcomes with respect to the unknown Werner parameters $\{w_i\}$ associated with different links~\cite{r21}. 
For a probability distribution $p_k$ of outcomes, the QFIM entry for parameters $w_i$, $w_j$ is defined as
\begin{equation}
F_{ij} = \sum_k \frac{1}{p_k} 
\left( \frac{\partial p_k}{\partial w_i} \right) 
\left( \frac{\partial p_k}{\partial w_j} \right),
\end{equation}
where the sum runs over all measurement outcomes. 
In the following, we list closed-form Fisher information expressions for the three measurement schemes.

\subsubsection{Estimation using LZM}
For a directly monitored link $i\in E$, the Fisher information contribution is
\begin{equation}
FL_{ii}^{(0)} = \frac{2}{(1+w_i)(1-w_i)}.
\end{equation}
If a monitor is not located at either end node of a given link, that link must be monitored indirectly through another monitor along a multi-link path. For such a path $P$ that involves multiple links, the Fisher information includes cross-coupling terms between the link parameters, expressed as
\begin{equation}
FL_{ij}^{(P)} = 
\frac{\prod_{\substack{{l \in P}\\l \ne i}} w_l \prod_{\substack{{l \in P}\\l \ne j}} w_l}{(1+\prod_{l \in P}w_l)(1-\prod_{l \in P} w_l)} .
\end{equation}

\subsubsection{Estimation using JBM}
Following the framework in \cite{r4}, for directly monitored link $i\in E$, we have
\begin{equation}
FJ_{ii}^{(0)} = \frac{12 w_i^2}{(1 + 3 w_i^2)(1 - w_i^2)}.
\end{equation}
For multi-link paths $P$ the cross-terms become
\begin{equation}
FJ_{ij}^{(P)} = 
\frac{12 w_i w_j}{(1 + 3 \prod_{l \in P} w_l^2)(1 - \prod_{l \in P}w_l^2)} 
\prod_{\substack{{l \in P}\\l \ne i}} w_l^2 \prod_{\substack{{l \in P}\\l \ne j}} w_l^2.
\end{equation}

\subsubsection{Estimation using PEM}
For the entanglement-assisted scheme, the Fisher information for directly monitored link $i\in E$ is
\begin{equation}
FP_{ii}^{(0)} = \frac{3}{(1 + 3 w_i)(1 - w_i)}.
\end{equation}
For indirect paths $P$, the cross-terms generalize to
\begin{equation}
FP_{ij}^{(P)} = 
\frac{3}{(1 + 3 \prod_{l \in P} w_l)(1 - \prod_{l \in P}w_l)} 
\prod_{\substack{{l \in P}\\l \ne i}} w_l \cdot \prod_{\substack{{l \in P} \\l \ne j}} w_l.
\end{equation}
PEM avoids the quadratic noise compounding of JBM.

\begin{figure*}[t]
\centering
\subfloat[\label{1}]{\includegraphics[scale=0.33]{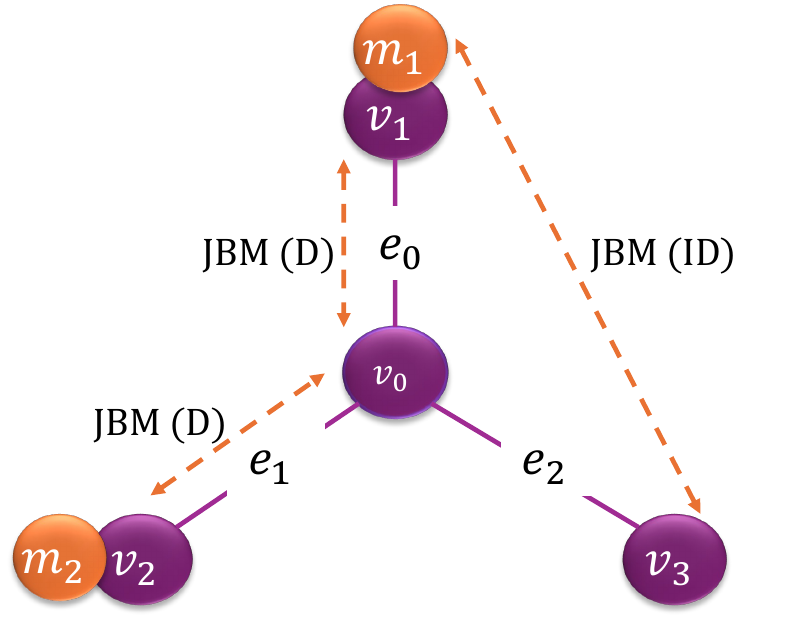}}%
\hfil
\subfloat[\label{2}]{\includegraphics[scale=0.33]{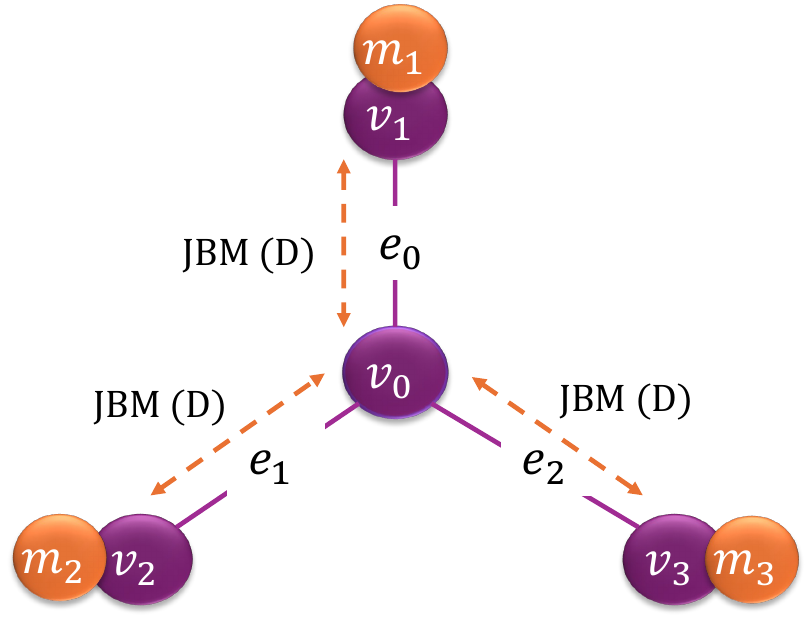}}%
\hfil
\subfloat[\label{3}]{\includegraphics[scale=0.33]{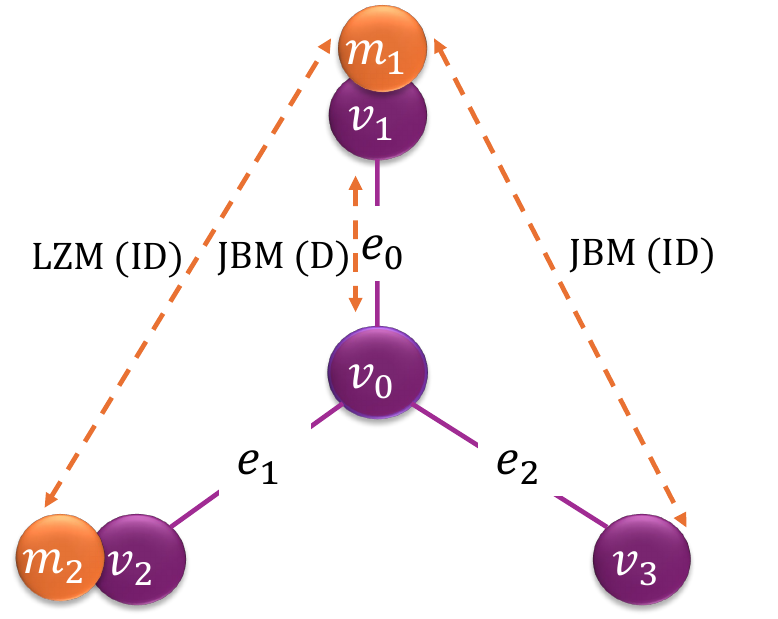}}%
\hfil
\subfloat[\label{4}]{\includegraphics[scale=0.33]{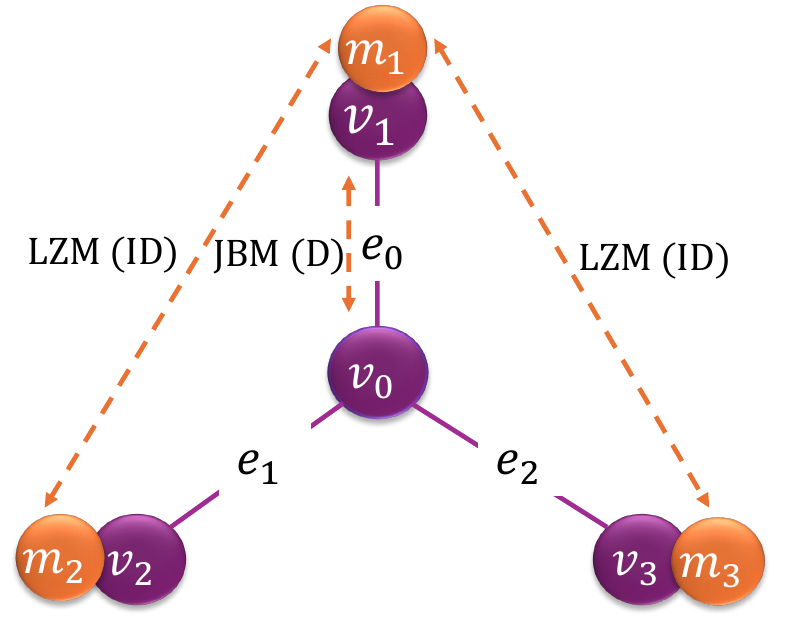}}%
\caption{Quantum Network Tomography of star networks. (a) two-monitor configuration using only JBM with two direct and one indirect JBM. (b) three-monitor configuration using only JBM with three direct JBMs. (c) two-monitor configuration combining JBM and LZM with one direct JBM, one indirect JBM, and one indirect LZM. (d) three-monitor configuration combining JBM and LZM with one direct JBM and two indirect LZMs.}
\label{fig2}
\end{figure*}

\subsection{QNT of Star Networks}

QNT attempts to estimate link parameters for every link in a network. The three state distribution and measurement schemes defined previously allow for the estimation of parameters characterizing network paths. We now describe two different strategies for estimating link parameters based on LZM and JBM for a four-node star-shaped quantum network with configurations involving two and three monitors. The two strategies do not utilize PEM since generating the required entanglement is resource intensive.

The JBM-only strategy, illustrated in Fig.~\ref{1} and Fig. \ref{2} for the two monitor and three monitor configurations, respectively, utilizes JBM exclusively to estimate link parameters. In Fig.~\ref{1}, two monitors are positioned at the end nodes $v_1$ and $v_2$. The corresponding links $e_0$ and $e_1$ are estimated through direct JBMs by the monitors $m_1$ and $m_2$, respectively. Link $e_2$ is estimated through indirect JBM by the monitor $m_1$, using the result for the link $e_0$ to isolate its contribution to the path $(e_0, e_2)$. In Fig.~\ref{2}, the monitors are located at the end nodes $v_1$, $v_2$, and $v_3$, allowing for direct JBM-based estimation of links $e_0$, $e_1$, and $e_2$.

The hybrid strategy, illustrated in Fig.~\ref{3} and Fig.~\ref{4} for the two and three monitor cases, respectively, combines JBM and LZM to estimate link parameters. In Fig.~\ref{3}, link $e_0$ is estimated by direct JBM, link $e_2$ is estimated by indirect JBM, and link $e_1$ is monitored through indirect LZM. In Fig.~\ref{4}, link $e_0$ is estimated directly via JBM, while links $e_1$ and $e_2$ are monitored through indirect LZM. In this strategy, the implementation of direct LZM is not feasible, as the monitor placement in the hub is excluded from consideration, as it was identified as the baseline for the end-to-end estimation in \cite{r4}.

Estimation is performed by generating multiple copies of the entangled states following the two defined strategies. For fair comparison, the QFIM values are normalized by the total number of channel uses for each strategy---eight and six channel uses for the two- and three-monitor configurations, respectively---as they require a different number of channel uses to obtain one copy of the required states for each path shown in Fig.~\ref{fig2}. However, the pattern of individual channel utilization differs between strategies. The first strategy employs link $e_0$ four times and links $e_1$ and $e_2$ twice each in the two-monitor case, whereas it uses each of the links $e_0$, $e_1$ and $e_2$ twice in the three-monitor case. The second strategy uses $e_0$ five times, $e_1$ once, and $e_2$ twice in the two-monitor case, while it employs $e_0$ four times and uses $e_1$ and $e_2$ once each in the three-monitor case.

% In this case, the total number of channel uses required to generate one copy of each path shown in Fig.~\ref{fig2} is the same across all strategies and equal to eight. Albeit, individual channels are used differently. The first strategy utilizes $e_0$ four times, and $e_1$ and $e_2$ two times. The second strategy utilizes $e_0$ five times, $e_1$ one time and $e_2$ two times.

\section{Numerical Results}

\begin{figure*}[t]
\centering
\subfloat[\label{5}]{\includegraphics[scale=0.38]{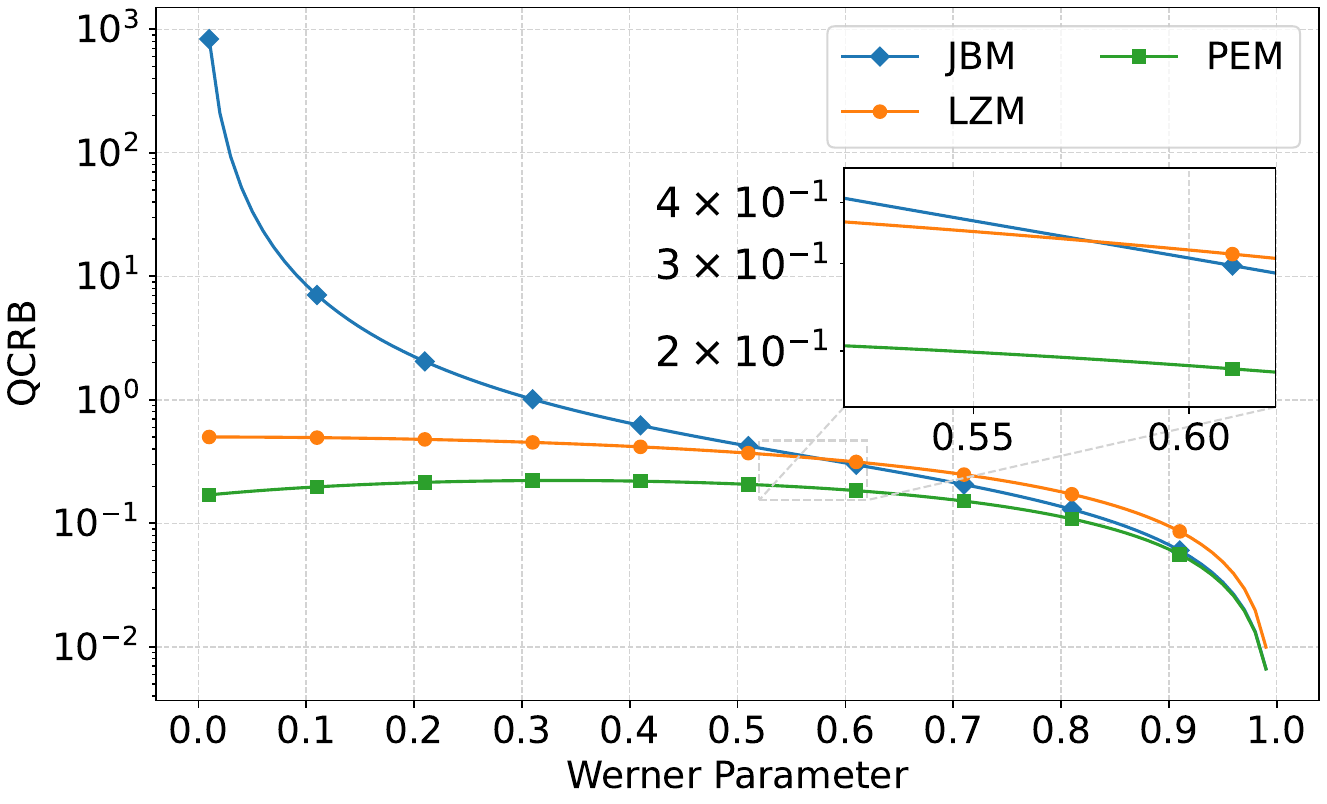}}%
\hfil
\subfloat[\label{6}]{\includegraphics[scale=0.35]{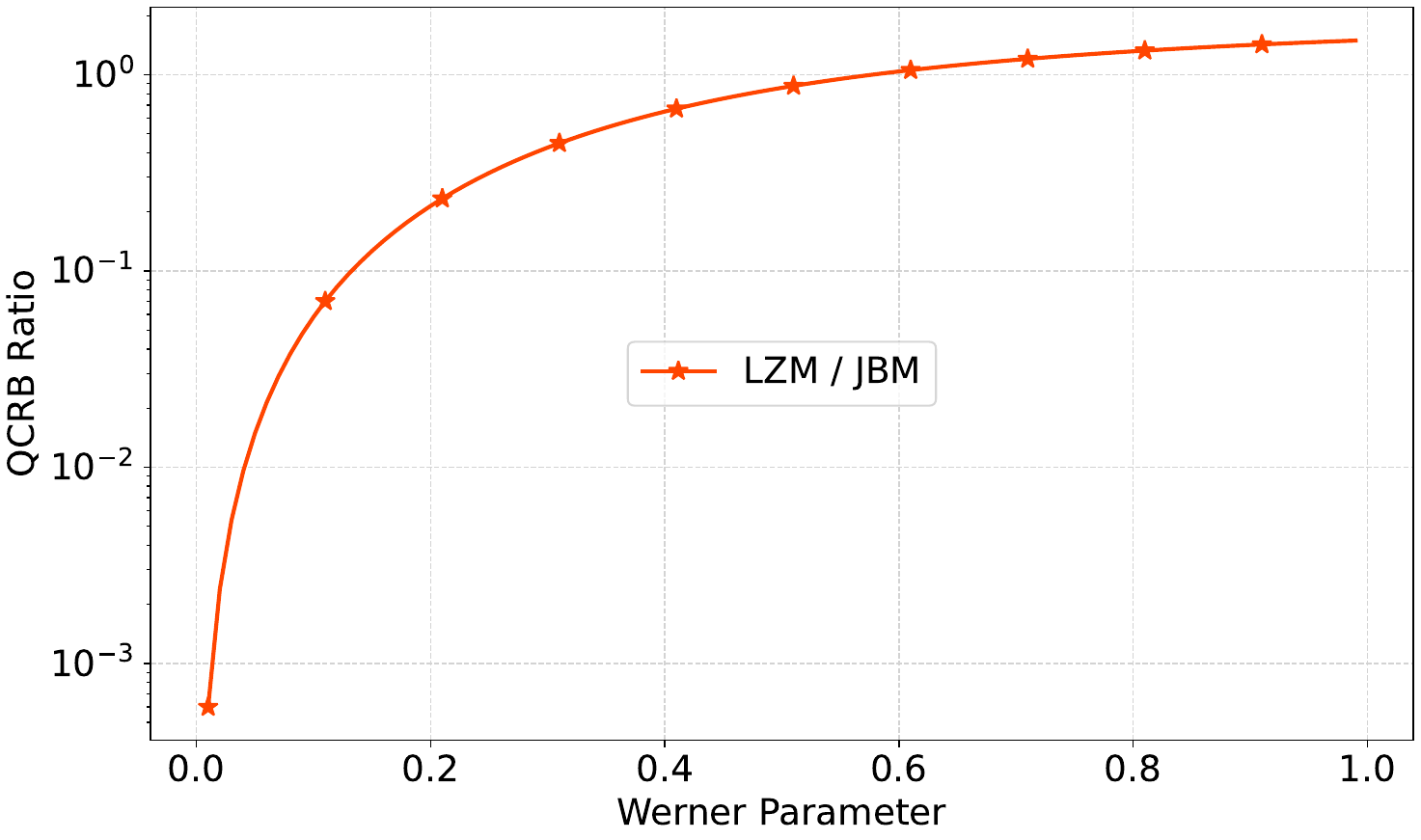}}%
\caption{Numerical QCRB analysis of a single-link network. (a) Variation of the QCRB with the Werner parameter $w$ for LZM, JBM, and PEM. The inset enlarges the $w$ region where the curves overlap. (b) Ratio between LZM and JBM across varying Werner parameter values.}
\end{figure*}

\begin{figure*}[t]
\centering
\subfloat[\label{7}]{\includegraphics[scale=0.38]{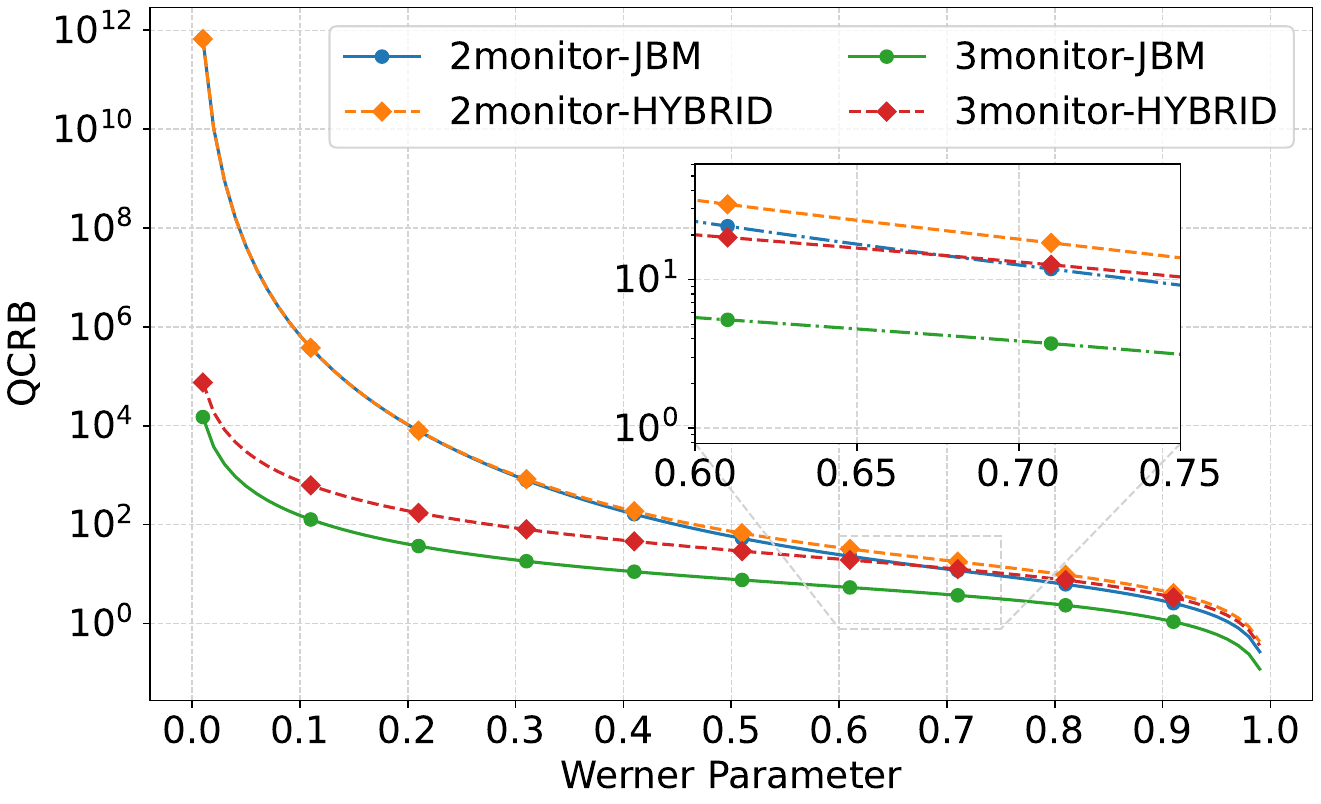}}%
\hfil
\subfloat[\label{8}]{\includegraphics[scale=0.38]{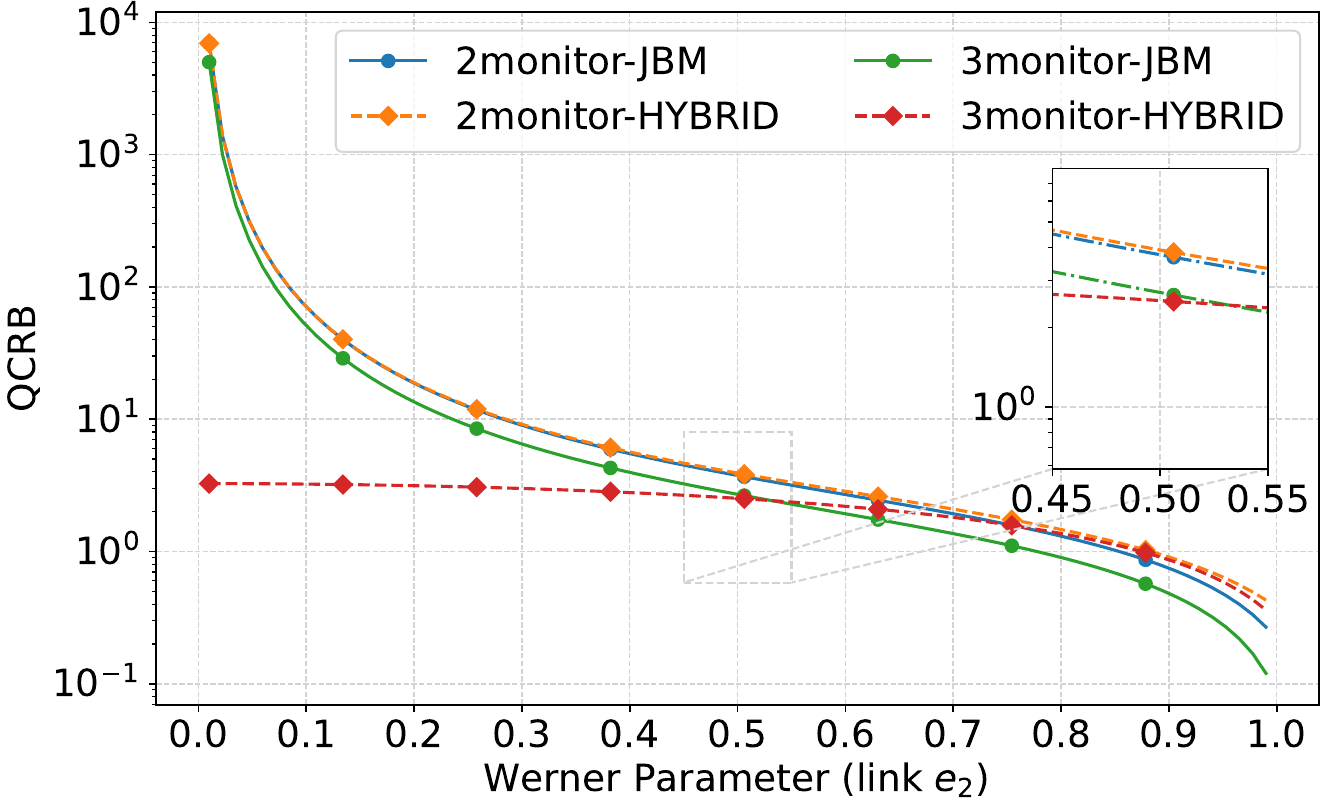}}%
\caption{Numerical QCRB analysis for a four-node star network involving two and three monitors. (a) JBM-only and hybrid (JBM+LZM) configurations with $w_1, w_2, w_3$ varying from 0.01 to 0.99. (b) JBM and hybrid configurations with fixed $w_0 = w_1 = 0.99$ and $w_2$ varying from 0.01 to 0.99.}
\label{9}
\end{figure*}

This section presents a numerical evaluation of the three proposed measurement schemes under different network configurations. The analysis first considers a single-link scenario, followed by a four-node quantum star network, where monitors are placed at two and three end nodes. The estimation performance of each scheme is compared in terms of the QCRB~\cite{r21}.

Fig.~\ref{5} presents the variation of the QCRB as a function of the Werner parameter $w$ for single-link direct-measurements across the three proposed schemes. The results indicate that the QCRB decreases monotonically with increasing 
$w$, as higher Werner parameters correspond to states with stronger entanglement and reduced depolarizing noise, leading to higher Fisher information and, consequently, smaller estimation variance. PEM achieves the lowest QCRB among the three methods, demonstrating the highest estimation precision. Moreover, JBM and LZM exhibit distinct trends: JBM achieves lower QCRB values for higher $w$ when compared to LZM, although its performance degrades more rapidly as $w$ decreases and LZM performs better than JBM in the low-$w$ regime ($w \leq 0.577)$.

This behavior is further explored in Fig.~\ref{6}, which presents the ratio of the QCRB between LZM and JBM, plotted as a function of the Werner parameter $w$. As $w$ increases, the ratio gradually approaches one, indicating that both schemes provide comparable precision in the high-fidelity regime, where the distributed states have fidelity close to one with the maximally entangled state $\ket{\Phi^{+}}$. This behavior stems from the fact that noise accumulates quadratically in JBM measurement statistics, while it accumulates linearly in LZM.

Fig. \ref{9} compares the QCRB for the four monitor configurations (illustrated in Fig. \ref{fig2}) as a function of the Werner parameter $w$. In the case of homogeneous links where $w_1$,$w_2$ and $w_3$ vary simultaneously (Fig.~\ref{7}), the configuration using JBM consistently achieves a lower QCRB than the hybrid strategy for both two and three-monitor networks. This indicates that JBM provides higher estimation precision under uniform noise conditions across all links.

In contrast, Fig.~\ref{8} shows the QCRB in the heterogeneous case where links $e_0$ and $e_1$ have fixed $w$ values and link $e_2$ experiences varying noise. Here, a similar pattern is observed for the two-monitor configuration, with JBM maintaining superior performance. However, in the three-monitor configuration, the hybrid approach outperforms JBM-only in the high-noise regime (lower $w$ values), and aligns with the trends reported in the single-link analysis shown in Fig. \ref{5}.

\section{Conclusion}
The results presented in this work provide a concrete foundation for selecting suitable measurement strategies in practical quantum network tomography. Three distinct state distribution and measurement schemes were analyzed: LZM, JBM, and PEM. Closed-form expressions for the QFIM were derived for each scheme, revealing how estimation precision scales with the Werner parameter, which characterizes the depolarizing nature of quantum channels. 

Numerical evaluations show that the PEM scheme achieves the lowest QCRB, providing the highest estimation accuracy. The JBM method offers a favorable balance between estimation performance and resource cost, as it requires only a single monitor per path. In contrast, the LZM approach demands two monitors located at the end nodes of each path, which increases resource usage but simplifies the measurement process. While LZM generally offers lower estimation precision than the other methods, it outperforms JBM in high-noise regimes for single-link estimation. Similarly, in high-noise conditions with heterogeneous links in a four-node star network, the hybrid approach combining LZM and JBM surpasses the JBM-only strategy, making it a practical choice in scenarios prioritizing fewer BSMs.

Future work will extend this study by developing an optimization framework that dynamically selects the most suitable state-distribution and measurement scheme to enhance estimation accuracy. In addition, realistic monitor placement models will be developed to account for non-idealities such as entanglement generation cost and operational errors, thereby enabling scalable, adaptive, and hardware-aware quantum channel characterization, facilitating the practical implementation of QNT.

\section*{Acknowledgment}

This research was supported by SFI (Science Foundation Ireland) grant 21/US-C2C/3750 for CoQREATE (Convergent Quantum REsearch Alliance in Telecommunications), CONNECT-2 grant 13/RC/2077-P2, and by the NSF-ERC Center for Quantum Networks grant EEC- 1941583.

\end{document}